\documentclass[sigconf]{acmart}
\AtBeginDocument{%
  
}

\copyrightyear{2025}
\acmYear{2025}
\setcopyright{acmlicensed}
\acmConference[CVMP '25]{European Conference on Visual Media Production}{December 3--4, 2025}{London, United Kingdom}
\acmBooktitle{European Conference on Visual Media Production (CVMP '25), December 3--4, 2025, London, United Kingdom}
\acmDOI{10.1145/3756863.3769713}
\acmISBN{979-8-4007-2117-5/2025/12}

\acmSubmissionID{13}

\usepackage{tabularx}
\usepackage{makecell}
\usepackage{balance}
\usepackage{setspace}

\begin{document}

\title{A Fast Volumetric Capture and Reconstruction Pipeline for Dynamic Point Clouds and Gaussian Splats}

\author{Athanasios Charisoudis}
\email{athanasios.charisoudis@epfl.ch}
\orcid{0000-0003-4769-7813}
\affiliation{%
  \institution{École Polytechnique Fédérale de Lausanne (EPFL)}
  \city{Lausanne}
  \country{Switzerland}
}
\affiliation{%
  \institution{Lucerne University of Applied Sciences and Arts}
  \city{Lucerne}
  \country{Switzerland}
}

\author{Simone Croci}
\email{simone.croci@hslu.ch}
\orcid{0000-0002-2979-804X}
\affiliation{%
  \institution{Lucerne University of Applied Sciences and Arts}
  \city{Lucerne}
  \country{Switzerland}
}

\author{Lam Kit Yung}
\email{kityung.lam@hslu.ch}
\orcid{0000-0003-2616-5463}
\affiliation{%
  \institution{Lucerne University of Applied Sciences and Arts}
  \city{Lucerne}
  \country{Switzerland}
}

\author{Pascal Frossard}
\email{pascal.frossard@epfl.ch}
\orcid{0000-0002-4010-714X}
\affiliation{%
  \institution{École Polytechnique Fédérale de Lausanne (EPFL)}
  \city{Lausanne}
  \country{Switzerland}
}

\author{Aljosa Smolic}
\email{aljosa.smolic@hslu.ch}
\orcid{0000-0001-7033-3335}
\affiliation{%
  \institution{Lucerne University of Applied Sciences and Arts}
  \city{Lucerne}
  \country{Switzerland} 
}

\renewcommand{\shortauthors}{Charisoudis et al.}

\begin{abstract}
We present a fast and efficient volumetric capture and reconstruction system that processes either RGB-D or RGB-only input to generate 3D representations in the form of point clouds and Gaussian splats. For Gaussian splat reconstructions, we took the GPS-Gaussian regressor and improved it, enabling high-quality reconstructions with minimal overhead. The system is designed for easy setup and deployment, supporting in-the-wild operation under uncontrolled illumination and arbitrary backgrounds, as well as flexible camera configurations, including sparse setups,  arbitrary camera numbers and baselines. Captured data can be exported in standard formats such as PLY, MPEG V-PCC, and SPLAT, and visualized through a web-based viewer or Unity/Unreal plugins. A live on-location preview of both input and reconstruction is available at 5–10 FPS. We present qualitative findings focused on deployability and targeted ablations. The complete framework is open-source, facilitating reproducibility and further research: \textit{{\color{blue}\url{https://github.com/irc-hslu/capturestudio}}}
\vspace{-1mm}
\end{abstract}

\begin{teaserfigure}
  \includegraphics[width=0.96\textwidth]{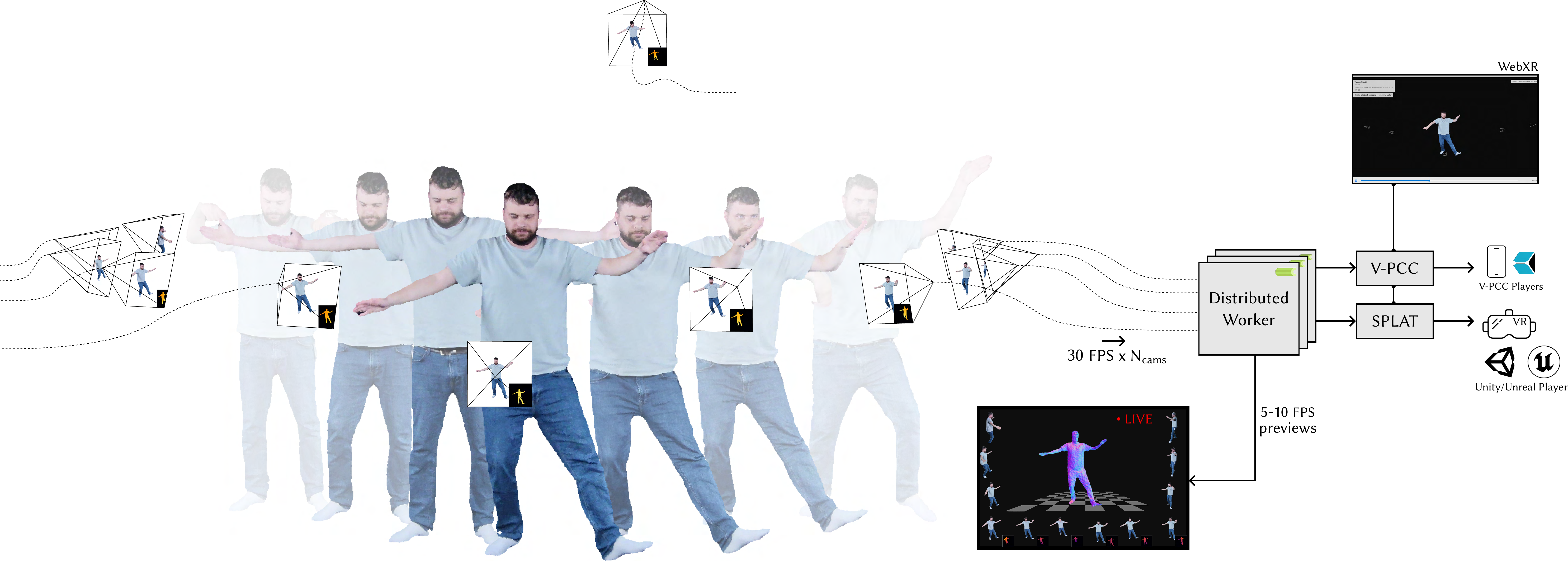}
  \caption{Teaser figure showing the proposed volumetric video capture and reconstruction system on a 12-camera setup.}
  \Description{}
  \label{fig:teaser}
\end{teaserfigure}


\maketitle

\section{Introduction}

Volumetric capture promises to carry live performance from stages into immersive media, telepresence, and interactive experiences. Yet going from capture to volumetric video still requires expensive and difficult-to-deploy multi-camera rigs, controlled studio conditions, and heavyweight post-processing \cite{humanrf, neuraldome, renderme, 4dgs, gpsgaussian}. Meanwhile, volumetric systems are increasingly deployed in research and industry, yet streamlined and flexible pipelines that bridge capture, reconstruction, and delivery remain scarce. We therefore present a fast and efficient volumetric capture and reconstruction system that is easy to deploy, supports relatively inexpensive cameras arranged in arbitrary configurations, operates under in-the-wild conditions, and produces volumetric reconstructions suitable for on-location preview and integration with external tools.

In particular, our system handles both RGB-D and RGB-only capture: when depth is present, it performs depth to color alignment and joint bilateral filtering; when depth is absent, it recovers it via calibrated stereo, allowing the system to run with commodity RGB cameras. Moreover, our solution is flexible enough to support any camera configuration, including sparse and irregular camera layouts with variations in camera count, placement, and baseline length; for our experiments we use 6–12 Orbbec Femto Bolt cameras arranged along a semicircular arc facing a small stage, at a mean radius of 2.5 m. 

The supported outputs are conventional point clouds suitable for standards-compliant interchange and compression \cite{ply, vpcc}, and also innovative Gaussian splats, which enable efficient rendering and high-fidelity reconstructions \cite{3dgs, threejs_gs, supersplat}. The developed pipeline derives both representations from a single, shared preprocessing stream.

The proposed pipeline starts with synchronized multi-view acquisition. Then, we run a unified preprocessing stage comprising performer segmentation, depth to color alignment, optical flow estimation, and image/flow–guided joint bilateral depth filtering on the GPU. From this shared intermediate point, we simultaneously produce (i) per-camera point-cloud reconstructions exported to PLY \cite{ply} and compressed with MPEG V-PCC \cite{vpcc}, and (ii) per-camera Gaussian splats, exported to PLY and packaged in SPLAT. These output formats are supported by existing players \cite{antimatter15_splat, supersplat, threejs_gs}, game engines and various tools. Moreover, we provide a web-based viewer that can display both representations and with WebXR support \cite{webxr}. In addition, we develop plugins for dynamic Gaussian splat playback in Unity and Unreal Engine 5.

For the generation of Gaussians, we chose to use GPS-Gaussian \cite{gpsgaussian}, a feed-forward regressor that converts RGB-D to pixel-wise Gaussians. We assessed that it does not perform well with the tested camera configurations, so we implement some improvements, namely, the introduction of Gaussian rotation re-parameterization and supervised fine-tuning. 

The developed system is modular, throughput-oriented, and standards-compliant. New depth estimators and reconstruction back ends can be added without changing the dataflow. 

Our evaluation concentrates on deployability; we present qualitative dynamic reconstruction results and side-by-side visual comparisons across representations and along the processing stages. We also include targeted ablations that isolate the effect of improvements we made to the chosen Gaussian regressor.

The contributions can be summarized as follows.
\vspace{-1mm}
\begin{itemize}
    \item We present a fast and efficient volumetric capture and reconstruction system that works with RGB-D or RGB-only input and produces point clouds and Gaussian splats; it is designed for fast setup and in-the-wild conditions, supporting any camera configuration. The system also provides on-location preview at 5–10\,FPS.
    \item We export reconstructions to PLY, MPEG~V-PCC, and SPLAT for compatibility with existing standards, and we build a web-based viewer with WebXR support plus plugins for dynamic Gaussian splat playback in Unity and Unreal.
    \item We improve the GPS-Gaussian regressor by introducing the Gaussian rotation re-parameterization, and fine-tuning on in-domain data, yielding more stable and higher-quality reconstructions across camera configurations.
    \item We evaluate the proposed volumetric system with dynamic reconstructions, visual comparisons, and ablations.
    \item We release the codebase of the pipeline, web-viewer and game engines plugins, along with representative samples.
\end{itemize}

\section{Background}

\begin{figure*}
    \centering
    \includegraphics[width=0.95\linewidth]{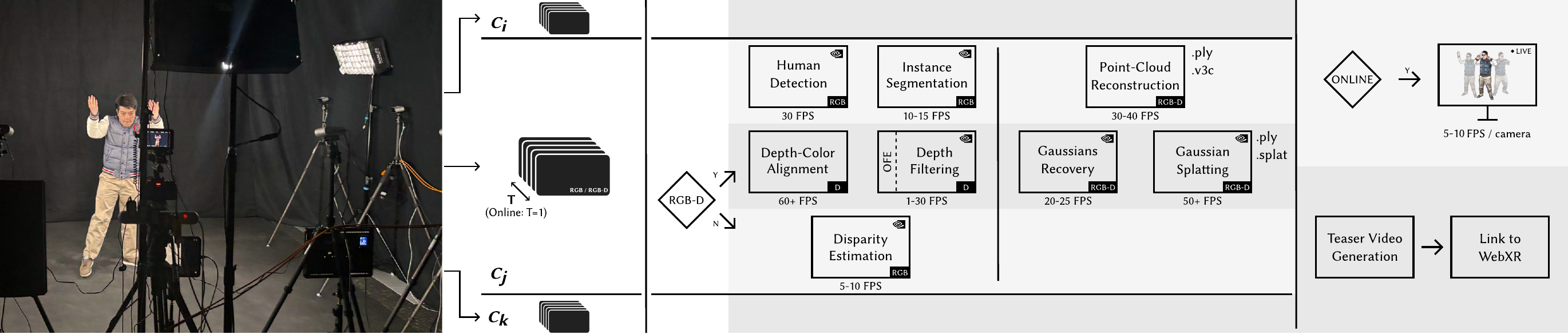}
    \caption{Left: capture in action. Right: pipeline overview (data flow from left to right). Per-camera RGB-D/RGB inputs are processed in parallel; vertical bars indicate synchronization points. Segmentation and depth processing run in parallel, then point-cloud and Gaussian-splat reconstructions are computed in parallel. A live monitor shows reconstruction previews at 5 to 10 FPS, cycling across input cameras; after processing, teaser clips and reconstructions are served in the web viewer.}
    \vspace{-3mm}
    \label{fig:pipeline}
\end{figure*}

Volumetric capture and reconstruction systems exist in both academia and industry, with academic work emphasizing methodological innovation and reproducibility, and commercial solutions focusing on scalable, robust end-to-end pipelines; this section reviews selected examples from both domains.

Among the academic solutions, \cite{Collet2015} presents a volumetric video pipeline that reconstructs dynamic scenes as textured meshes from multi-camera input and enables efficient compression and streaming for interactive free-viewpoint rendering. 
LiveScan3D \cite{kowalski2015livescan3d, livescan3d} is an open-source system for real-time 3D reconstruction using multiple Kinect sensors, generating colored point clouds and supporting streaming to Unity and HoloLens for immersive mixed-reality applications. 
Holoportation \cite{orts2016holoportation} is a real-time 3D capture technology developed by Microsoft Research that reconstructs, compresses, and transmits high-quality 3D models of people, enabling remote users to see, hear, and interact with each other in mixed reality environments. 
VolumetricCapture \cite{sterzentsenko2018low, volumetricapture} is a flexible, low-cost, and portable multi-sensor framework that facilitates real-time 3D reconstruction using commodity hardware, supporting applications such as free-viewpoint video and immersive telepresence. 
The system presented in \cite{schreer2019capture} demonstrates a full volumetric video pipeline, including multi-camera acquisition, geometry and texture reconstruction, and spatio-temporal compression for efficient storage and playback.
EasyVolcap \cite{xu2023easyvolcap} is an open-source PyTorch library designed to accelerate neural volumetric video research, providing a unified pipeline for multi-view data processing, 4D scene reconstruction, and dynamic volumetric video rendering.
ImViD pipeline \cite{yang2025imvid} uses 46 synchronized cameras to capture 5K multi-view video at 60 FPS, enabling high-fidelity volumetric reconstruction for immersive 6-DoF VR experiences. ImViD also integrates spatially aligned audio captured alongside multi-view video, enabling immersive sound rendering.

Of the commercially available solutions, Depthkit Studio \cite{depthkitstudio} is a portable volumetric video system that enables full-body 360° 3D capture using multiple RGB-D cameras, facilitating real-time streaming and integration into Unity for immersive applications.
Evercoast \cite{evercoast} is a company that provides an end-to-end volumetric video solution enabling the transformation of multi-camera video into high-fidelity 4D spatial data for immersive applications. Their system supports scalable capture setups, ranging from compact to large-scale installations.
ScannedReality \cite{scannedreality} offers a streamlined volumetric video solution that utilizes a mobile multi-camera rig, enabling real-time 3D point cloud previews, mesh reconstruction, and seamless deployment across Unity, Unreal Engine, and web platforms.
HOLOSYS \cite{4dviews} by 4Dviews is a volumetric capture system designed for high-quality 3D video production. It provides seamless integration with Unity, Unreal Engine, and WebXR, facilitating real-time rendering and immersive experiences. Additionally, HOLOSYS+ allows users to choose between mesh-and-texture or Gaussian splat formats.
8i \cite{8i} is a company that offers end-to-end solutions for capturing, transforming, and streaming high-fidelity holograms across various platforms, including VR, AR, and web applications. Their proprietary system utilizes multi-camera capture stages, advanced machine learning algorithms, and real-time compression techniques.
Volu \cite{volograms} by Volograms is an AI-powered mobile application that enables users to create, transform, and share dynamic 3D volumetric holograms using just a smartphone, facilitating immersive AR experiences with features like full-body filters and collaborative sharing.
Volucap \cite{volucap} is a company specializing in high-resolution volumetric video capture, offering solutions, which utilizes a 42-camera system to produce over 3,000 megapixels per frame, enabling the creation of lifelike 3D human models for applications in virtual reality, augmented reality, and digital media production.
Microsoft’s Mixed Reality Capture Studios \cite{microsoft} provide a professional volumetric video system that employs dense multi-camera rigs and advanced reconstruction pipelines to generate high-quality, production-ready holographic content.

Compared to the volumetric systems presented before, the proposed volumetric capture and reconstruction system is an academic open-source solution. 
It can take as input RGB-D or RGB-only and output reconstructions in two possible representations, namely, classical point clouds and innovative Gaussian splats. 
The system is fast and efficient, offering easy setup and in-the-wild operation, while supporting arbitrary camera configurations, including sparse setups.
For playback, we additionally provide a web-based viewer and game engine plugins, together with an on location preview.

\section{System Overview}

The proposed pipeline receives synchronized multi-view imagery from RGB-D or RGB-only rigs, execute a unified preprocessing stack comprising performer segmentation, depth–color alignment, optical-flow estimation, and image- and flow-guided bilateral filtering on the GPU, and then branch to point-cloud and Gaussian-splat reconstruction backends. The design targets human performance capture with sparse to moderately dense arrays (in the order of 10 cameras) rapid setups in unprepared environments, and therefore operate with low-cost, portable hardware. It should provide low-latency previews suitable for on-site decision making while preserving an offline mode for more expensive processing and improved fidelity. Furthermore, it should scale with the number of sensors and available hardware without reconfiguration, and remain robust to variation in sensor placement and baseline lengths as well as intermittent device or node failures. Outputs support standards-compliant formats, including PLY and MPEG V-PCC for point clouds and PLY and SPLAT for Gaussian splats, while a web-based viewer enables inspection of dynamic sequences. Typical deployments involve compact stages, commodity frame rates and resolutions, and unconstrained illumination and backgrounds. In offline mode, the same preprocessing admits higher-quality variants, including optional color correction and enhanced depth filtering.

\subsection{Capture Setup}

We employ 6–12 Orbbec Femto Bolt RGB–D sensors arranged along a semicircular arc facing a compact stage area. The camera selection was based on their programmability and depth quality; in qualitative comparisons against two Intel RealSense models and the Azure Kinect DK we observed more stable depth and improved color fidelity. The stage footprint measures approximately \(4\times 4\,\mathrm{m}\), with performers operating near the arc's center at distances of \(0.5\)–\(3.5\,\mathrm{m}\), and adjacent cameras were spaced at roughly \(0.9\,\mathrm{m}\) baselines. Each device provides time-of-flight (ToF) depth via an infrared emitter/receiver pair, and to mitigate mutual interference we schedule the emitters in a round-robin pattern with a \(140\,\mu\mathrm{s}\) offset. Camera clocks are synchronized to the host at connection time and re-synchronized every five minutes, enabling timestamp-based alignment in both online and offline processing modes, while hardware trigger and time distribution are supplied via Orbbec SyncHub Pro switches over Ethernet. We terminate a session if inter-camera timestamp drift exceeds \(17\,\mathrm{ms}\), ensuring synchronized data recording.

All cameras are mounted rigidly and monitored via onboard inertial measurements to detect micro-motion. In practice, we discard sessions in which the measured angular acceleration exceeds \(0.1\,\mathrm{rad}/\mathrm{s}^2\) or the linear acceleration exceeds \(10^{-3}\,\mathrm{m}/\mathrm{s}^2\). Devices are centrally configured from a workstation in a star topology, where a multi-process producer–consumer architecture initializes sensors, enforces manual focus, exposure, and white balance, and orchestrates frame acquisition and storage. Unless noted otherwise, we record color at \(3840\times 2160\) and \(30\,\mathrm{FPS}\), and depth at \(640\times 576\) and \(30\,\mathrm{FPS}\); typical settings include an exposure of \(120\,\mu\mathrm{s}\) and a white balance of \(3700\,\mathrm{K}\), with scene-dependent contrast reduction under harsh illumination. Data are buffered in batches, synchronized by timestamps prior to persistence, written in HDF5 containers to local SSD, and mirrored to a network-attached storage array.

Camera calibration techniques are employed to compute the intrinsic and extrinsic parameters for all cameras. For this, we use printed checkerboard and ChArUco patterns that are swept through individual frusta and across multi-view intersections, after which pattern detections in the recorded frames are processed by standard OpenCV-based routines followed by bundle adjustment to recover per-camera parameter and pose matrices. We evaluate two popular open-source libraries, Caliscope~\cite{caliscope} and MultiCamCalib~\cite{multicamcalib}, and adopt the latter owing to its diagnostics and curation tools, which enable detection and removal of high-residual images prior to optimization. When in offline mode, color response is equalized using color-checker targets and three-dimensional LUTs derived in DaVinci Resolve. We use the ColorChecker Video by X-Rite in our experiments. For depth-to-color alignment we rely on factory calibration of the infrared (IR) sensors, which remains stable under controlled temperature conditions used in our indoor captures.

\subsection{Pre-Processing Pipeline}
The processing pipeline is based on distributed task queues and central orchestration. We use Celery~\cite{celery} to define the task graph, and maintain separate queues for the CPU and GPU. At the beginning of each capturing session, we spawn workers for both queues, the number of which depends on the system capability. In our case, we use 16 CPU and 2 GPU workers, as most of our experiments take place in a single-GPU machine. The task nodes and a schematic of the pipeline blueprint is given in Figure \ref{fig:pipeline}. To enable robustness, we follow an Extract-Transform-Load (ETL) approach for the node execution, i.e. each node reads its input and stores its output on the disk. Consequently, among the different workers it is only data paths that travel, allowing for task retries in case of failures, at the expense of slightly increased latency.

The pipeline supports online and offline execution modes. In either mode, data flow through mostly the same stages during pre-processing. To increase throughput we split the processing into two stages. During the first stage, each camera stream is processed independently and in parallel with the other streams enabling seamless scaling when more processing nodes or additional compute resources are available. In the second stage, we sync the data among the different cameras, disregarding the ones for which not all cameras were actively recording, and proceed with the dynamic reconstructions. During online mode, the data are already synced, so the syncing step is skipped; otherwise the frame timestamps are used to cluster the data into multi-view frames.

\subsubsection{RGB Processing}

\begin{figure}
    \centering
    \includegraphics[width=0.9\linewidth]{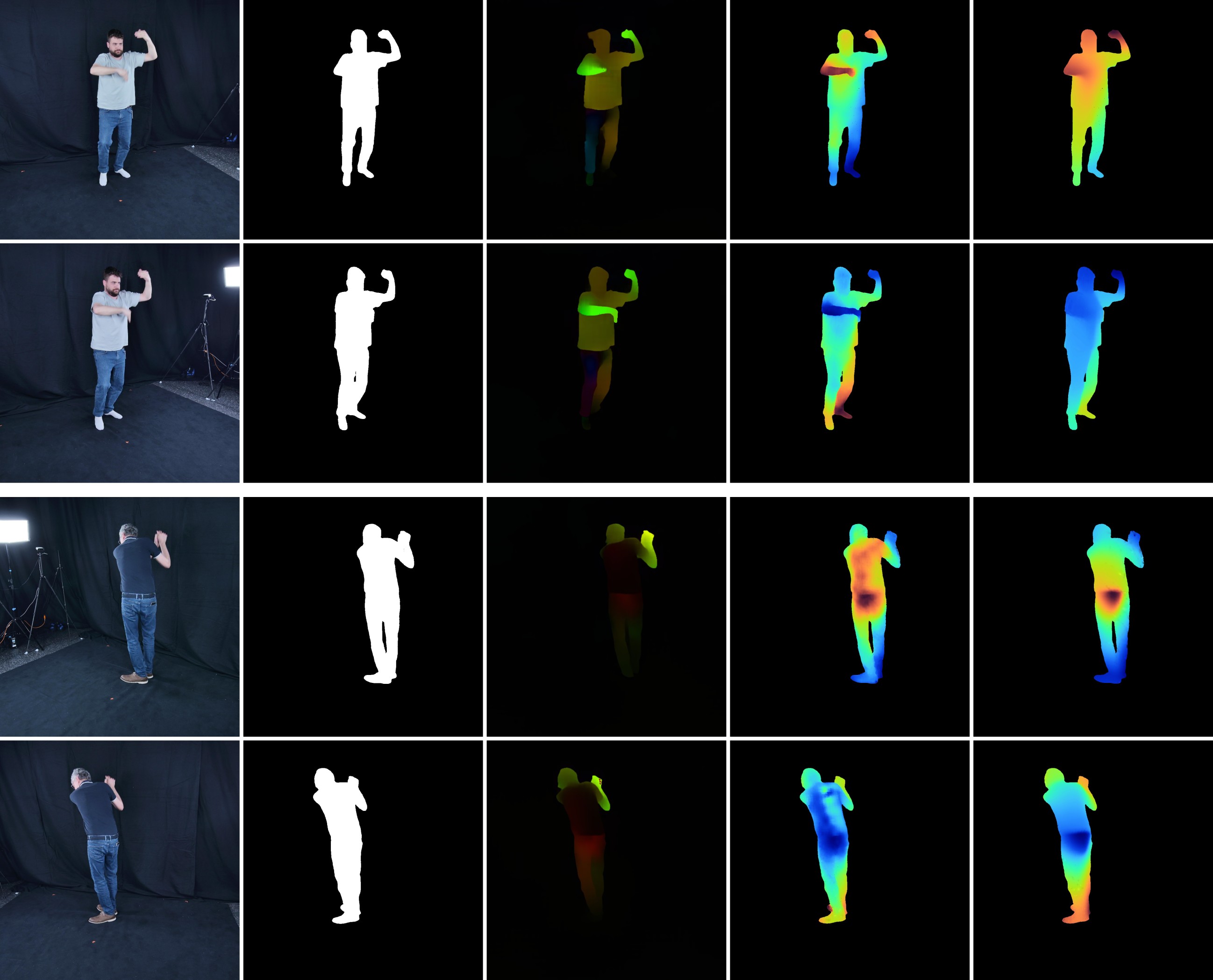}
    \caption{Color cues for two sample image pairs. From left to right: color, mask, optical flow, disparity from RAFT-Stereo \cite{raftstereo} and from FoundationStereo \cite{foundationstereo}. RAFT-Stereo was trained on human data only, predicting more accurate disparity ranges.}
    \label{fig:color_processing}
    \vspace{-4mm}
\end{figure}

Initially, the RGB cues are fed to modern human detection and instance segmentation models. After initialization the models remain on the GPU facilitating efficient resource usage; we follow this practice with all deep networks employed in our pipeline. For detecting humans in the initial frames we use YOLO 11 \cite{yolo11}, followed by SegmentAnythingV2 (SAM2) \cite{sam2}, the state-of-art instance segmentation model at the time of writing. During online mode, we feed the images one-by-one to the segmentor, while in offline mode, a video is created from the frames of each camera. In both cases, the segmentation masks are computed using the video mode of SAM2, which  provides temporally consistent human masks and also tracking across frames. This enables us to create human tracks, where in each track we have the color and corresponding mask for each frame in time. In our setup, SAM2 runs at 10-15 FPS when fed batches of 2 images on an Nvidia RTX 4080 GPU. Given more GPUs, our design allows spawning correspondingly more GPU workers for an approximately analogous speedup in masks generation, thus allowing seamless scaling up. 

During offline mode, in parallel to the instance segmentation, we estimate the apparent motion of the pixels for each color sensor independently, which we exploit during subsequent depth filtering tasks. For this, we employ VideoFlow \cite{videoflow} the state-of-art multi-frame optical flow estimation (OFE) model at the time of writing of this paper. It uses 3 frames, estimating the optical flow from the two edge ones towards the middle one. Using VideoFlow's 3-frame prediction network, this node operates at 1-2FPS, making it the bottleneck of our entire processing pipeline and forbidding integration in the online mode. It is important to note here, that we tested other OFE methods, including RAFT \cite{teed2020raft} and FlowFormer \cite{flowformer} which we found inferior. In practice, during the online mode we defer the optical flow computation and depth processing that relies on it for post-capture, to avoid increased latencies unless it is explicitly requested to do else-wise. Figure \ref{fig:color_processing} shows the outputs of the processing steps that each color frame undergoes.

\subsubsection{Depth Processing}
Given our choice of inexpensive and portable depth sensors, the captured depth cues are usually noisy. This includes normals noise, holes due to sensor errors (phase quantization, IR reflective surfaces etc), and also IR-crosstalk especially when multiple depth sensors are used simultaneously. Consequently, depth cleaning and filtering is vital for having high fidelity reconstructions. In our case depth signal is aligned to color either during or after capturing, where the latter enables more sophisticated interpolation kernels to be used if needed. In both cases, misalignment issues usually remain, especially when the captured depth is at VGA resolution, 6x smaller than the RGB resolution. As a consequence, data denoising is an essential first step to our reconstruction algorithms that rely on sensor depth cues.

 To get rid of misaligned depth pixels that are far from the human foreground, we use a quantile-based outlier removal. In particular, we keep all the points inside the foreground that fall in the 99.9th quantile. As a result only misaligned points that belong to the background and when the background is significantly far from the subject (in the order of 1 m or more) are removed. Furthermore, we use a Canny edge erosion on the foreground masks, and invalidate all removed pixels in the depth maps. This helps us minimize the normal noise which is usually concentrated around occlusion boundaries (Figure \ref{fig:depth_filtering} left).
 
 Finally, we perform guided bilateral filtering~\cite{tomasi1998bilateral} on the aligned yet noisy depth maps. Two such variants were developed. The first uses the color signal as guidance essentially implementing the following equation:
\[
D_{f}(x) =
\frac{
\displaystyle \sum_{x_i \in \Omega_x} D(x_i) \;
G_{\sigma_s}(\|x_i - x\|) \;
G_{\sigma_r}(\|I(x_i) - I(x)\|)
}{
\displaystyle \sum_{x_i \in \Omega_x}
G_{\sigma_s}(\|x_i - x\|) \;
G_{\sigma_r}(\|I(x_i) - I(x)\|)
}
\]
\noindent
where $D$ is the noisy depth map and $D_f$ the filtered one, $x$ and $x_i$ denote 2D vectors on the image plane, $\Omega_x$ is the neighborhood window centered at $x$ including only valid foreground depth pixels, $G_{\sigma}$ is the Gaussian kernel $G_{\sigma}(x) = \frac{1}{\sigma \sqrt{2 \pi}} \; \text{exp}(-\frac{x^2}{2 \sigma^2})$ where $\sigma$ is the standard deviation, $G_{\sigma_s}$ and $G_{\sigma_r}$ are the spatial and range kernels, respectively, and $I$ is the RGB signal. 
When optical flows are available, we use a newly developed filtering approach, in an attempt to minimize remaining high-frequency depth noise and fill the holes. We first warp the filtered depth $D^{t-1}_{f}$ and the image $I^{t-1}$ at frame $t-1$ to the current frame $t$ by using the computed (backward) optical flow field obtaining $D^{t-1 \rightarrow t}_f(x)$ and $I^{t-1 \rightarrow t}(x)$, respectively. Then, we filter according to the following equation:

\begin{align*}
D^t_{f}(x) = &
\frac{
\displaystyle \sum_{x_i \in \Omega_x} D^t(x_i) \;
G_{\sigma_s}(\|x_i - x\|) \;
G_{\sigma_r}(\|I^t(x_i) - I^t(x)\|)\; + 
}{
\displaystyle \sum_{x_i \in \Omega_x}
G_{\sigma_s}(\|x_i - x\|) \;
G_{\sigma_r}(\|I^t(x_i) - I^t(x)\|)\; +
} \\
& \frac{\lambda_t \; D^{t-1 \rightarrow t}_f(x) \; G_{\sigma_t}(\|I^{t-1 \rightarrow t}(x) - I^t(x)\|)}{\lambda_t \; G_{\sigma_t}(\|I^{t-1 \rightarrow t}(x) - I^t(x)\|)}
\end{align*}
\noindent
where $\sigma_t$ is the temporal standard deviation, and $\lambda_t$ is a weight that controls the importance of the temporal component of the filter. 

\begin{figure}
    \centering
    \includegraphics[width=0.80\linewidth]{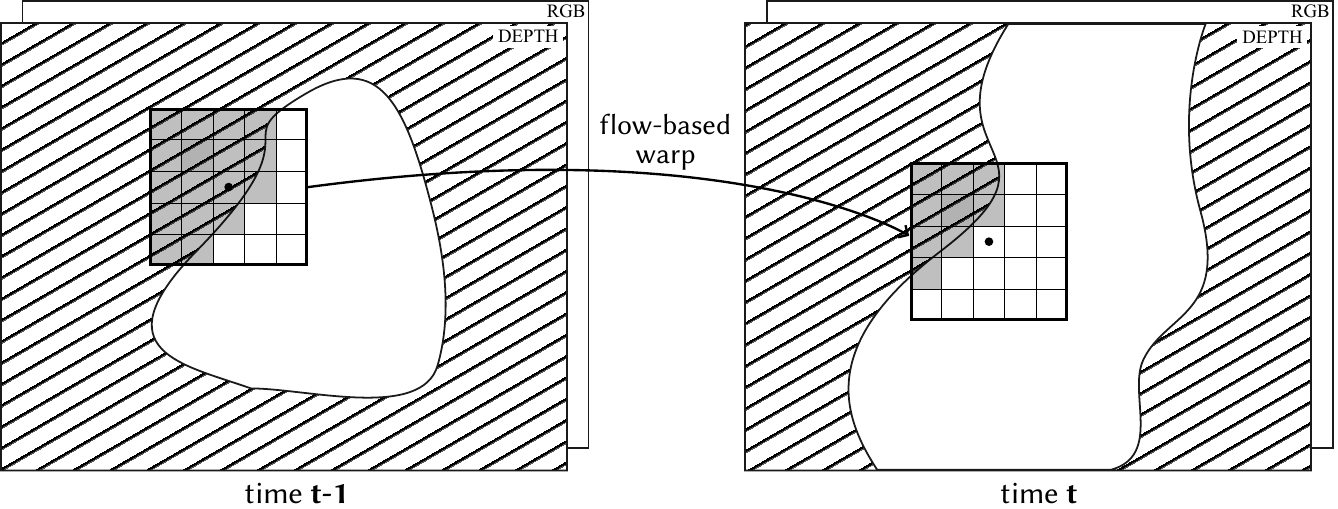}
    \caption{Illustration of how the bilateral spatiotemporal filter is applied to two consecutive frames. The stripes deonote valid depth values, while the white regions denote holes.}
    \vspace{-2mm}
    \label{fig:bilateral_filter}
\end{figure}

\begin{figure}
    \centering
    \includegraphics[width=0.85\linewidth]{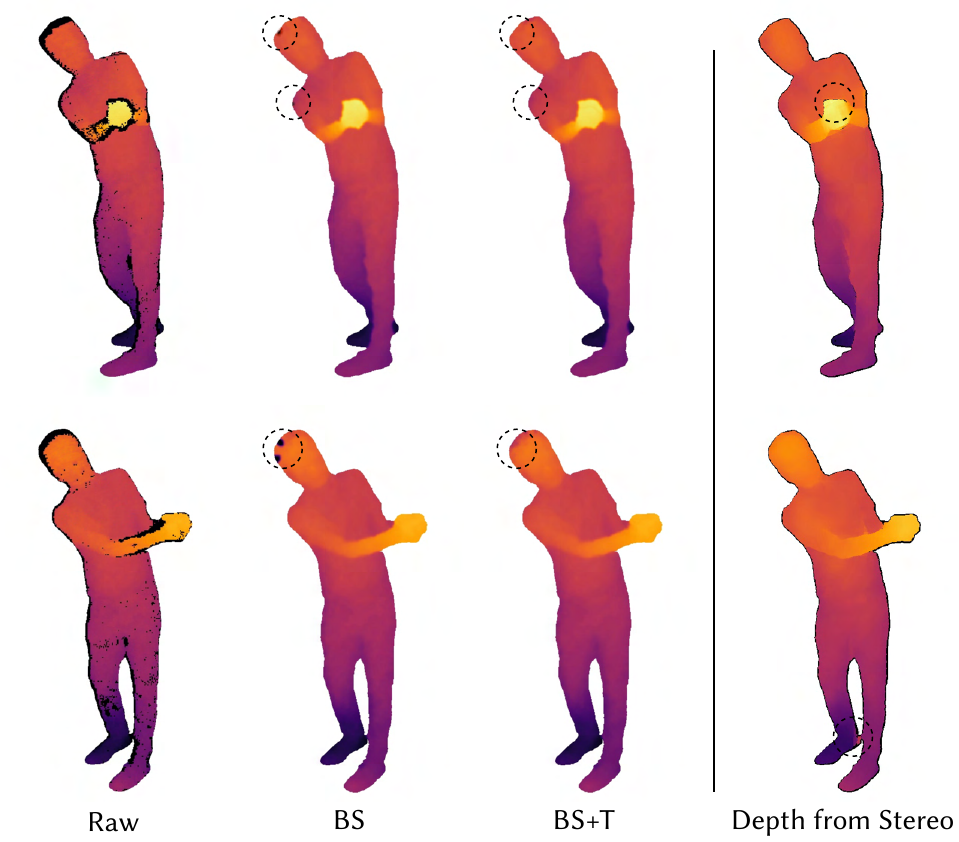}
    \caption{Sensor depth (left block): raw, spatially filtered (BS), and spatio-temporally filtered (BS+T). Stereo-estimated depth (right block) is computed from rectified pairs and shown without bilateral filtering.}
    \vspace{-2mm}
    \label{fig:depth_filtering}
\end{figure}

We call this filter bilateral spatiotemporal (BS+T) as opposed to the bilateral spatial (BS) for the non-temporal version. Using custom CUDA kernels, both filter veresions run at over 60 FPS on an RTX 4080. An illustration of how the BS+T is applied to two consecutive RGB-D frames is given in Figure \ref{fig:bilateral_filter}. In our experiments, we work with 4K (aligned) depth signals, and use a square window $\Omega_x$ of radius $r{=}7$ pixels (i.e., $15{\times}15$), spatial kernel $\sigma_s{=}7$ pixels, range kernel $\sigma_r{=}0.1$ (RGB units on a 0–1 scale with $\ell_2$ color differences), temporal kernel $\sigma_t{=}0.06$, and weight $\lambda_t{=}0.6$ for the BS+T variant; these fixed settings provide stable edge preservation with limited flicker across all the tested sequences. Figure \ref{fig:depth_filtering} (left block) shows a comparison on real data before and after depth filtering.

\subsubsection{Depth from Color}
\label{subsec:depth_from_stereo}
Additionally, the pipeline supports estimating depth from color, which both mitigates sensor–depth failure modes and enables RGB–only operation in environments where active infrared sensing is unreliable. When reliable sensor depth is available this step is skipped.

We compute depth by rectified stereo. For an adjacent camera pair with focal length \(f\) (pixels) and baseline \(B\), we rectify the images, infer a disparity field \(d(u,v)\), and convert disparity to depth via
\[
Z(u,v)=\frac{f\,B}{d(u,v)}.
\]
Modern disparity estimation networks are typically trained to predict \emph{positive} disparities for a fixed input ordering (e.g., left–right), which complicates inference when the geometric ordering is reversed. We therefore employ a flipping identity that preserves positivity without retraining. Let \(\mathrm{flip}(\cdot)\) denote a horizontal mirror and \(d_{L\!\to\!R}\) the disparity predicted on inputs \((L,R)\). For rectified pairs of width \(W\),
\[
d_{R\!\to\!L}(u,v)
\;=\;
-\,d_{L\!\to\!R}(W-1-u,v)
\;=\;
-\;\mathrm{flip}\!\left(d_{L\!\to\!R}\right)(u,v),
\]
so that
\[
-\,\mathrm{flip}\!\left(d(L,R)\right)\;=\;d\!\left(\mathrm{flip}(L),\,\mathrm{flip}(R)\right).
\]
This identity restores the positive–disparity convention for either input order, after which \(Z=fB/d\) applies unchanged. We apply the flipping patch whenever the median \(x\)-coordinate of the foreground in the first image is smaller than that of the second: both inputs are mirrored horizontally, and the resulting disparity is mirrored and negated to enforce the positive-disparity convention. 

We evaluate two disparity estimators. \emph{FoundationStereo} \cite{foundationstereo}, a state-of-the-art model trained on generic imagery, produces tends to over-smooth human geometry and to merge adjacent limbs with the torso, which hinders reconstruction. \emph{RAFT-Stereo} \cite{raftstereo} is retrained following GPS-Gaussian on synthetic THuman2.1 \cite{thuman} renders, except that we generate imagery and ground-truth depth using virtual cameras that match our actual setup rather than the circular layouts used in the default training. This model preserves body part separation and yields satisfactory metric depth in our captures, as can be seen on samples in Figure \ref{fig:depth_filtering} (right). As with all rectified stereo, applicability degrades as the baseline widens beyond the common field of view; we therefore restrict stereo to adjacent pairs whose overlap remains substantial.

\section{Dynamic Reconstruction}

\begin{figure*}
    \centering
    \includegraphics[width=0.9\linewidth]{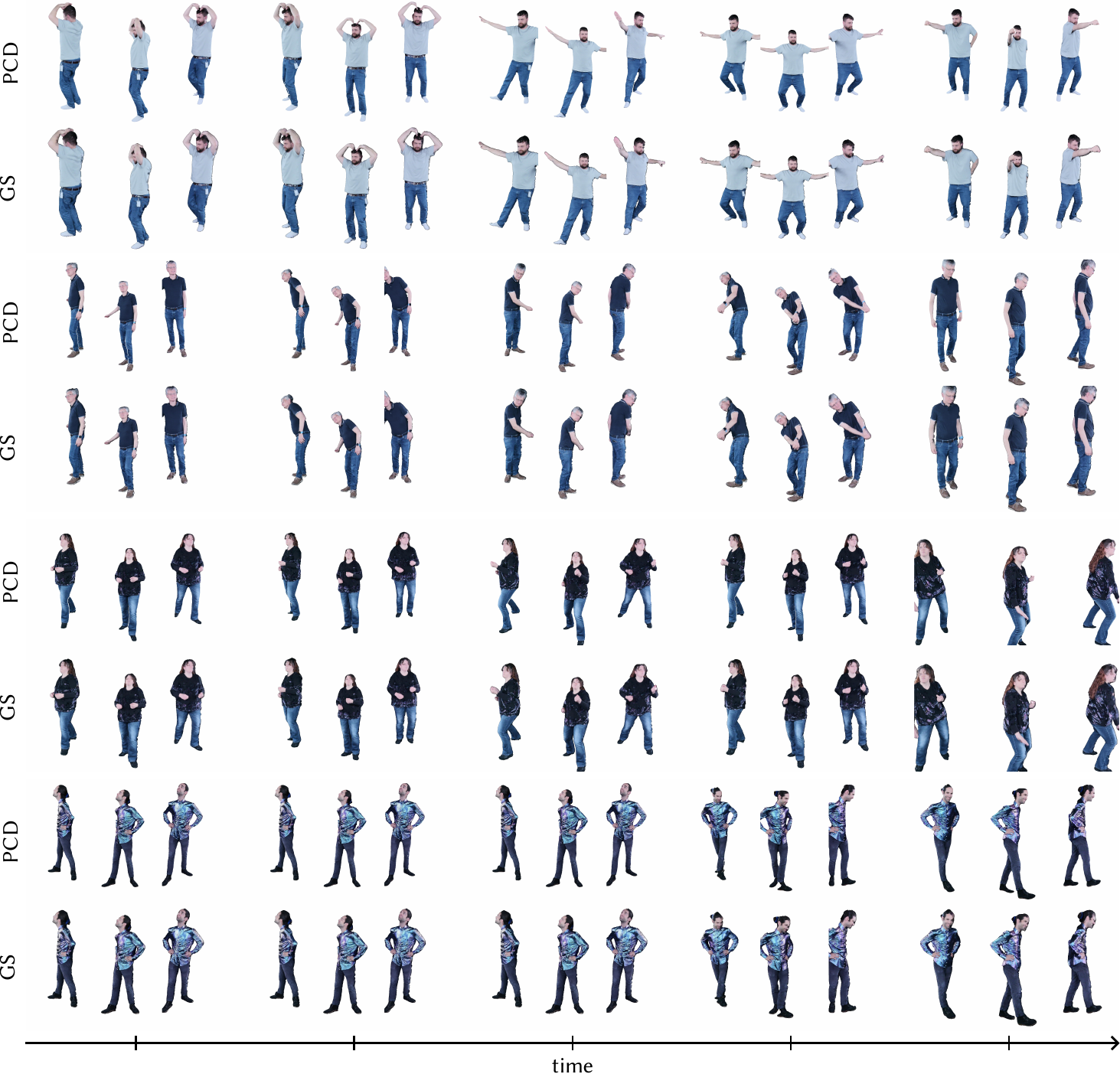}
    \caption{Sample outputs of the dynamic reconstruction pipeline. We randomly select five timesteps for four subjects from the output of the pipeline. For each subject point clouds and Gaussian splat reconstructions are given in triplets, while we visualize three views per triplet.}
    \vspace{-2mm}
    \label{fig:pcd_vs_gs}
\end{figure*}

From segmented RGB images and aligned depth maps, the pipeline reconstructs a time‐ordered sequence of 3D representations without temporal tracking or multi-view fusion, a choice that enables predictable latency and straightforward parallelization in either online or offline modes. Two reconstruction branches run in parallel (Figure \ref{fig:pipeline} middle). The first generates per-camera point clouds that are exported, streamed to the live monitor, and compressed for archival and distribution. The second generates per-camera Gaussian splats whose attributes are predicted from image and depth. The web viewer selects, at each virtual viewpoint, the source camera whose point of view is closest to the virtual camera and switches representations accordingly.

\subsection{Point Clouds}
Given a calibrated camera \(i\) with intrinsic matrix \(K_i\) and pose matrix \(_\mathrm{c}^\mathrm{w}T_i\) (adopting Craig's convention \cite{craig} and writing \({}_{\text{a}}^{\text{b}}\!T\) for the transformation that maps vectors from coordinate frame \(\text{a}\) to \(\text{b}\)), we back-project each foreground pixel \((u,v)\) with depth \(d(u,v)\) to a world-space 3D point and attach its color \(\mathbf{c}_i(u,v)\), thereby producing a colored point cloud per frame and per camera. For a pixel \((u,v)\) with depth \(d\) (meters), the camera-space point is
\[
^{\mathrm{c_i}}\mathbf{x}(u,v,d)=
\begin{bmatrix}
\frac{(u-c_x)}{f_x}\,d\\[2pt]
\frac{(v-c_y)}{f_y}\,d\\[2pt]
d
\end{bmatrix},
\quad
K_i=\begin{bmatrix}f_x&0&c_x\\0&f_y&c_y\\0&0&1\end{bmatrix},
\]
and the world-space point is \(^{\mathrm{w}}\mathbf{x}(u,v,d)={_\mathrm{c_i}^\mathrm{w}}R\,{^{\mathrm{c_i}}\mathbf{x}}(u,v,d)+{^\mathrm{w}}\mathbf{t}_{i_{ORG}}\) with \(_\mathrm{c}^\mathrm{w}T_{i}=\begin{bmatrix}{_\mathrm{c_i}^\mathrm{w}}R&{^\mathrm{w}}\mathbf{t}_{i_{ORG}}\\ \mathbf{0}^\top&1\end{bmatrix}\) being the pose matrix of camera \(i\).

To suppress residual background leakage and depth speckle, we apply a radius-based validity test in world space. For each point \(\mathbf{p}\), we count neighbors within radius \(r\) and keep \(\mathbf{p}\) only if \(|\{\mathbf{q}:\lVert \mathbf{q}-\mathbf{p}\rVert_2<r\}|\ge N_{\min}\). We use \(r=20\,\mathrm{cm}\) and \(N_{\min}=30\), evaluated efficiently with a radius-search K-dimensional tree \cite{kdtree, friedman1977algorithm}, which removes isolated outliers while preserving limb detail at typical working distances.

In RGB-only operation, each adjacent camera pair provides two rectified views and therefore two disparity maps related by the flipping identity described in Subsection \ref{subsec:depth_from_stereo}. We compute both left–right and right–left disparities, convert each to depth via \(Z=fB/d\), and associate the two resulting depth maps back to their originating cameras. Consequently, a physical rig of \(N\) cameras yields \(2(N-1)\) depth maps and point clouds per frame. All point clouds are written to disk for archival and compression.

Visualization uses a consistent camera-selection policy for both the live monitor and the web viewer. The virtual camera follows a cubic Bézier path that passes near the physical rig. In sensor-depth mode, the displayed cloud is the one acquired by the physical camera closest to the current virtual viewpoint. In stereo mode, the viewer switches at the virtual position nearest to the current physical camera, transitioning from the previous–current to the current–next stereo cloud so that the displayed geometry remains representative of the local viewpoint.

During online mode, per-camera clouds are serialized to PLY and streamed at interactive rates. We render with a fixed point size to maximize rasterization throughput and to match the projected sampling density at the operating range; at \(1024\times 1024\) display resolution and \(2.2\)–\(2.7\,\mathrm{m}\) viewing distances, a constant size of approximately two pixels yields near hole-free coverage without excessive overdraw. A median filter in the depth buffer mitigates quantization artifacts introduced by fixed-size splats. For geometric inspection, we estimate per-point normals by covariance analysis in local neighborhoods: neighbors within a \(0.1\,\mathrm{m}\) radius (up to \(30\) points) define a covariance whose minor-axis eigenvector provides the normal direction, yielding stable estimates at commodity sampling densities. In the teaser (Figure~\ref{fig:teaser}), we render a sample reconstruction shaded with per-point normals estimated via the covariance-based method described above.

For archival and distribution, the offline path encodes each per-camera sequence with MPEG V-PCC. We use the reference encoder with common test condition defaults, intra-only groups of pictures for short clips, and default patch packing. In our sequences with approximately \(90{,}000\)–\(120{,}000\) points per frame, we observe compression ratios between \(3\times\) and \(7\times\) depending on scene content; for example, \(300\) frames occupying roughly \(870\,\mathrm{MB}\) in PLY compress to about \(180\,\mathrm{MB}\). The resulting bitstreams play back in standards-compliant V-PCC players on desktop and mobile devices. In Table~\ref{tab:pcd_metrics} (PCD columns) we report median end-to-end latency and compression for point-cloud reconstructions under the 6- and 8-camera rigs. The online path meets the 5–10\,FPS preview target, and the offline encoder achieves \(\approx5{\times}\) size reduction with the reference V\textendash PCC settings for the selected sequences. Figure~6 (top rows) provides the corresponding qualitative results, showing per-frame point clouds for five subjects at matched times and viewpoints.

\begin{table}[t]
\caption{End-to-end latency and compression for dynamic reconstructions on real captures. 
Latency is the median per-frame time from sensor input to reconstructed asset emission in the online path (lower is better). 
Compression is the ratio of encoded size of a camera sequence to raw input (higher is better), using MPEG~V\textendash PCC for point clouds and \texttt{SPLAT} for Gaussian splats.}
\label{tab:pcd_metrics}
\centering
\begin{tabularx}{\columnwidth}{l *{4}{>{\centering\arraybackslash}X}}
\toprule
& \multicolumn{2}{c}{6 cameras} & \multicolumn{2}{c}{8 cameras} \\
\cmidrule(lr){2-3}\cmidrule(lr){4-5}
Metric & PCD (V\textendash PCC) & GS (\texttt{SPLAT}) & PCD (V\textendash PCC) & GS (\texttt{SPLAT}) \\
\midrule
Latency [ms] $\downarrow$                 & 130 & 160 & 180 & 230 \\
Compression ratio [\(\times\)] $\uparrow$ & 5.2 & 3.2 & 5.1 & 3.2 \\
\bottomrule
\end{tabularx}
\vspace{-5mm}
\end{table}

\subsection{Gaussian Splats}

The second reconstruction branch represents each frame as a set of per-camera Gaussian primitives whose positions and colors derive from the corresponding point clouds, while opacity, scale, and rotation are predicted from image evidence and depth using a pixel-wise regression network. This design maintains the per-frame, per-camera granularity that supports low-latency online processing and straightforward parallelization, and it avoids multi-view merging, which we find to degrade in-between viewpoints for sparse rigs.

We parameterize each primitive \(i\) by a world-space mean \(\boldsymbol{\mu}_i \in \mathbb{R}^3\), a symmetric positive-definite covariance \(\boldsymbol{\Sigma}_i \in \mathbb{R}^{3\times 3}\), a color \(\mathbf{c}_i \in [0,1]^3\), and an opacity \(\alpha_i \in [0,1]\). Positions \(\boldsymbol{\mu}_i\) and colors \(\mathbf{c}_i\) come from the point cloud back-projection, while the regression network predicts opacity, a rotation \(R_i \in \mathrm{SO}(3)\), and axis-aligned scales \(\mathbf{s}_i=(s_{x},s_{y},s_{z})\) for each valid pixel. The covariance is assembled as \(\boldsymbol{\Sigma}_i = R_i \,\mathrm{diag}(\mathbf{s}_i^2)\, R_i^\top\) in world coordinates. For rendering, a camera with extrinsics \((R,\mathbf{t})\) and intrinsics \(K\) projects \(\boldsymbol{\mu}_i\) to the image plane and induces a screen-space elliptical footprint with covariance \(\boldsymbol{\Lambda}_i = J\,R\,\boldsymbol{\Sigma}_i\,R^\top J^\top\), where \(J\) is the Jacobian of the perspective projection evaluated at \(R(\boldsymbol{\mu}_i-\mathbf{t})\). Contributions at pixel centers are weighted by a Gaussian kernel with covariance \(\boldsymbol{\Lambda}_i\) and composited in approximate front-to-back order using pre-multiplied alphas.

\subsubsection{Gaussian Rotations Re-parameterization}
\begin{figure}
    \centering
    \includegraphics[width=0.61\linewidth]{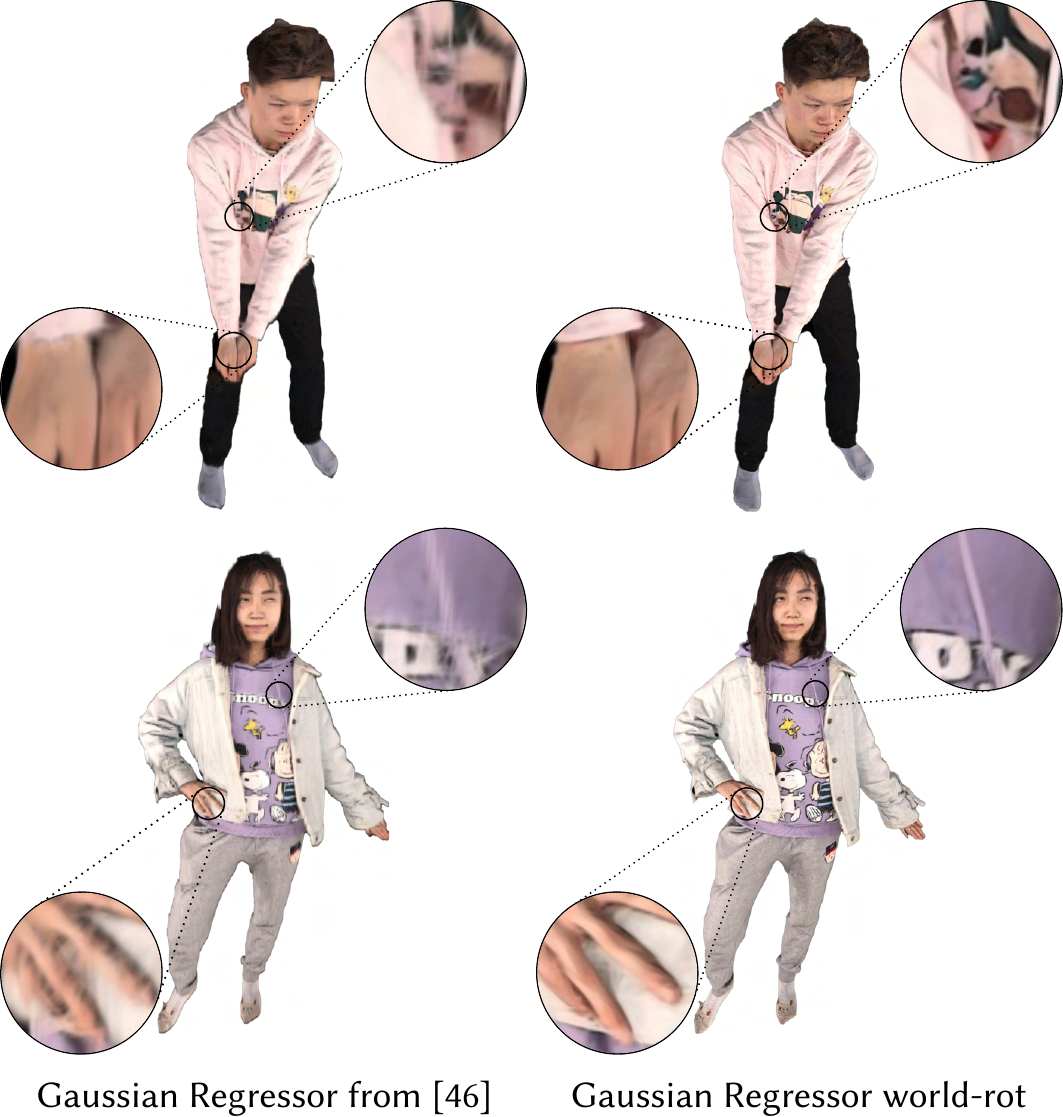}
    \caption{Validation results of the Gaussian regressor with and without Gaussian rotation re-parameterization.}
    \vspace{-2mm}
    \label{fig:gps_comparisons_thuman}
\end{figure}

During deployment we observed a recurring failure mode in pixel-wise splat regression that arose from a frame inconsistency: several implementations, including the GPS-Gaussian version we evaluated, predict rotations in the \emph{camera} frame and then use them as if they were expressed in \emph{world} coordinates. Circular, same-height camera layouts can hide this error by encouraging highly anisotropic “needle-like’’ splats that align with the training geometry; arbitrary rigs and non-coplanar camera sets expose the degeneracy as elongated, view–dependent artifacts. We correct this error by re-parameterizing the Gaussian rotations in the world frame. To isolate the effect of the rotation change, we train only the Gaussian regressor from \cite{gpsgaussian} and use ground-truth depth as input. Figure~\ref{fig:gps_comparisons_thuman} shows validation reconstructions before and after the correction.

Let \({}_{g}^{c}\!R\) denote the rotation of a Gaussian’s local frame \(g\) expressed in the \emph{camera} frame \(c\), as predicted by the regression network, and let \({}_{c}^{w}\!R\) be the calibrated camera-to-world rotation. The correct world-frame rotation of the Gaussian (given right-handed rotation systems) is then
\[
{}_{g}^{w}\!R \;=\; {}_{c}^{w}\!R\;{}_{g}^{c}\!R,
\]
i.e., the camera-to-world rotation must \emph{premultiply} the predicted camera-frame rotation.

With axis-aligned scales \(\mathbf{s}=(s_x,s_y,s_z)\) and \(S=\mathrm{diag}(\mathbf{s}^2)\), the world-frame covariance used by the rasterizer becomes
\[
{}^{w}\!\Sigma
\;=\;
{}_{g}^{w}\!R\, S \,\big({}_{g}^{w}\!R\big)^{\!\top}
\;=\;
{}_{c}^{w}\!R\;\Big({}_{g}^{c}\!R\, S \,\big({}_{g}^{c}\!R\big)^{\!\top}\Big)\;\big({}_{c}^{w}\!R\big)^{\!\top}.
\]
Equivalently, if one first forms the camera-frame covariance
\(\ {}^{c}\!\Sigma = {}_{g}^{c}\!R\, S \,\big({}_{g}^{c}\!R\big)^{\!\top}\), the change of frame is
\[
{}^{w}\!\Sigma \;=\; {}_{c}^{w}\!R\;{}^{c}\!\Sigma\;\big({}_{c}^{w}\!R\big)^{\!\top}.
\]
When the network also outputs a camera-frame Gaussian center \({}^{c}\!\boldsymbol{\mu}_{g}\), the corresponding world-space mean is
\({}^{w}\!\boldsymbol{\mu}_{g}={}_{c}^{w}\!R\;{}^{c}\!\boldsymbol{\mu}_{g} + {}^{w}\!\mathbf{t}_{c},\)
where \({}^{w}\!\mathbf{t}_{c}\) is the camera origin in world coordinates. In our system the mean position \({}^{w}\!\boldsymbol{\mu}_{g}\) comes from point-cloud back-projection and is already in the world frame; only the rotation, and therefore the covariance, requires re-parameterization.

Using \({}_{g}^{c}\!R\) directly as if it were a world-frame rotation, aligns Gaussian principal axes with the camera frame and produces viewpoint dependent elongations under non-circular layouts. Applying the transformation above removes this pathology. We retrain the Gaussian regressor with the same protocol as GPS-Gaussian, but replace camera-frame rotations by \({}_{g}^{w}\!R\) and feed ground-truth depth (skipping RAFT-Stereo module); all other settings remain unchanged. The patched model produces better-conditioned splats with improved sharpness and color fidelity. Quantitative results are reported in Table~\ref{tab:gps_rot_fix}, and qualitative examples in Figure~\ref{fig:gps_comparisons_thuman}, where the retrained regressor is denoted \textit{world-rot}.

\begin{table}[t]
\centering
\caption{Gaussian regressor before/after re-parameterization of Gaussian rotations.}
\label{tab:gps_rot_fix}
\begin{tabular}{lccc}
\toprule
Method & PSNR $\uparrow$ & SSIM $\uparrow$ & LPIPS $\downarrow$ \\
\midrule
GS-Regressor from ~\cite{gpsgaussian} & 36.13 & 0.947 & 0.049 \\
GS-Regressor world-rot & \textbf{37.69} & 0.964 & 0.048 \\
\bottomrule
\end{tabular}
\end{table}

\subsubsection{Finetuning of GPS-Gaussian}
We further align training with deployment by increasing synthetic inter-camera baselines and removing view stitching and multi-view loss aggregation while retaining world-space rotations; these changes reduce geometric bias and yield small improvements. Substantive gains arise from supervised fine-tuning (SFT) performed per camera configuration. 
We fine-tune on six performances captured with similar configurations, for a total of roughly \(72\mathrm{k}\) frames. The objective combines a photometric reconstruction term and a scale regularizer. The reconstruction loss averages over visible pixels \(\Omega\) in reprojected images and uses a weighted sum of \(\mathrm{SSIM}\) \cite{ssim} and the smooth-\(L_{1}\) (Huber) penalty,
\[
\mathcal{L}_{\mathrm{rec}}
=
\lambda_{\mathrm{L1}}\,
\frac{1}{|\Omega|}\!\sum_{p\in\Omega}
\phi_\delta\!\big(I(p)-\hat I(p)\big)
+
\lambda_{\mathrm{SSIM}}\,
\Big(1-\overline{\mathrm{SSIM}}(I,\hat I)\Big),
\]
with \(\lambda_{\mathrm{L1}}=0.8\) and \(\lambda_{\mathrm{SSIM}}=0.2\). The Huber penalty (with fixed \(\delta=0.05\)) is
\[
\phi(e)=
\begin{cases}
\dfrac{e^{2}}{2\times 0.05}, & |e|<0.05,\\[4pt]
|e|-\dfrac{0.05}{2}, & \text{otherwise}.
\end{cases}
\]
and \(\overline{\mathrm{SSIM}}(I,\hat I)\) denotes the mean SSIM computed over local \(11\times 11\) px windows,
\[
\mathrm{SSIM}(x,y)=
\frac{(2\mu_x\mu_y+C_1)(2\sigma_{xy}+C_2)}
{(\mu_x^2+\mu_y^2+C_1)(\sigma_x^2+\sigma_y^2+C_2)}.
\]

To prevent degenerate anisotropy in the predicted scales, we encourage a high-entropy distribution of scale magnitudes through a differentiable soft-histogram regularizer. Let \(\mathbf{s}_i=(s_{x,i},s_{y,i},s_{z,i})\) be the per-axis scales of Gaussian \(i\) and let \(\{b_m\}_{m=1}^{M}\) be uniformly spaced bin centers on \([0,s_{\max}]\). For axis \(a\in\{x,y,z\}\) we form a soft histogram
\[
h_{a,m}=\sum_{i}\exp\!\Big(-\tfrac{1}{2}\tfrac{(s_{a,i}-b_m)^2}{\sigma_h^2}\Big),
\qquad
p_{a,m}=\frac{h_{a,m}}{\sum_{k} h_{a,k}},
\]
compute the entropy \(H_a=-\sum_{m} p_{a,m}\log(p_{a,m}+\varepsilon)\), and minimize a clipped negative entropy,
\[
\mathcal{L}_{\mathrm{ent}}=\lambda_{\mathrm{ent}}\sum_{a}\max\!\big(0,\,H^\star-H_a\big),
\]
where \(H^\star\) is a target entropy threshold, \(\sigma_h\) is the kernel bandwidth, and \(\lambda_{\mathrm{ent}}=10^{-2}\). The full fine-tuning loss is
\[
\mathcal{L}=\mathcal{L}_{\mathrm{rec}}+\mathcal{L}_{\mathrm{ent}}.
\]

\begin{figure}
    \centering
    \includegraphics[width=0.95\linewidth]{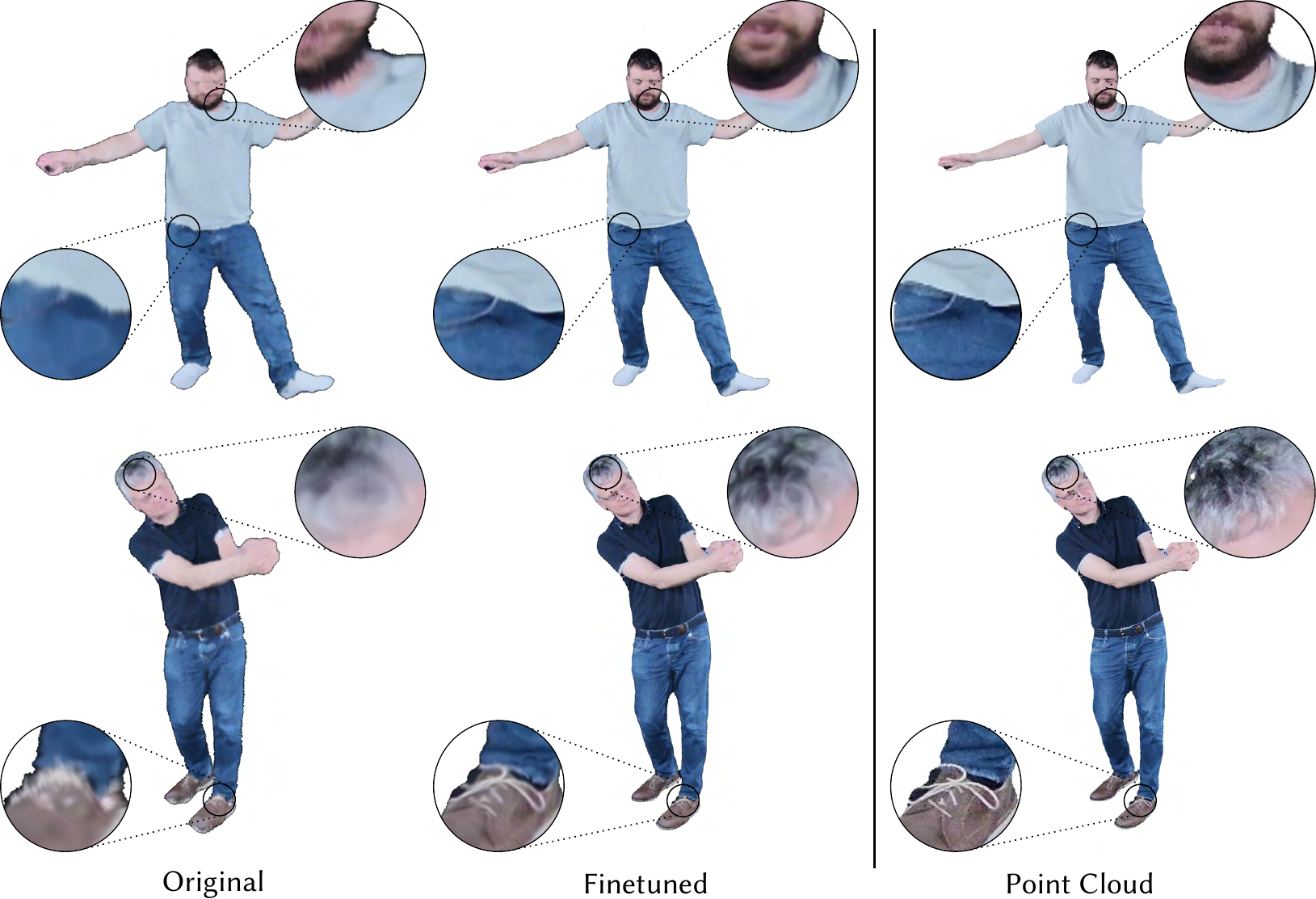}
    \caption{Visual comparisons of Gaussian splat reconstructions before and after fine-tuning, and point clouds.}
    \vspace{-2mm}
    \label{fig:gps_comparisons}
\end{figure}

We also modify the scale parameterization to improve transfer across camera distances and intrinsics. Instead of the softplus used in GPS-Gaussian, we first bound the pre-activations with a hyperbolic tangent and then apply per–axis instance normalization with learnable affine parameters,
\[
\hat z_{a,i}=\tanh(z_{a,i}),\qquad
\bar z_{a,i}=\frac{\hat z_{a,i}-\mu_{a}}{\sigma_{a}},\qquad
s_{a,i}=s_{\max}\big(\gamma_{a}\,\bar z_{a,i}+\beta_{a}\big),
\]
where \(a\in\{x,y,z\}\), \(\mu_{a}\) and \(\sigma_{a}\) are the per–axis instance statistics, and \(\gamma_{a},\beta_{a}\) are learned. The covariance uses squared scales \(S=\mathrm{diag}(s_{x,i}^2,s_{y,i}^2,s_{z,i}^2)\), which ensures positive definiteness while allowing the affine parameters to re-normalize scale magnitudes during fine-tuning. Empirically, the combination of the SSIM–Huber reconstruction loss and the entropy regularizer, together with the \(\tanh\)+instance-normalized scale parameterization, produces consistent improvements in robustness and perceived fidelity across venues and motions. Figure~\ref{fig:gps_comparisons} presents visual comparisons of Gaussian-splat reconstructions before and after fine-tuning and against the corresponding point-cloud reconstructions; despite these improvements, point clouds remain slightly sharper in most of our experiments.

\subsubsection{Rendering and Playback of Gaussian Splats}
We encode each frame’s splat data in a standalone \texttt{SPLAT} file and concatenate them into a video \texttt{SPLAT} stream (header with per–frame sizes followed by data blocks), following the spirit of \cite{4dgs1,4dgs2}. During packaging, colors and rotations are quantized to 8\,bits and scales to 16–32\,bits, which provides the compression reported in Table~\ref{tab:pcd_metrics} (SPLAT columns).

\begin{figure}[h]
    \centering
    \includegraphics[width=0.9\linewidth]{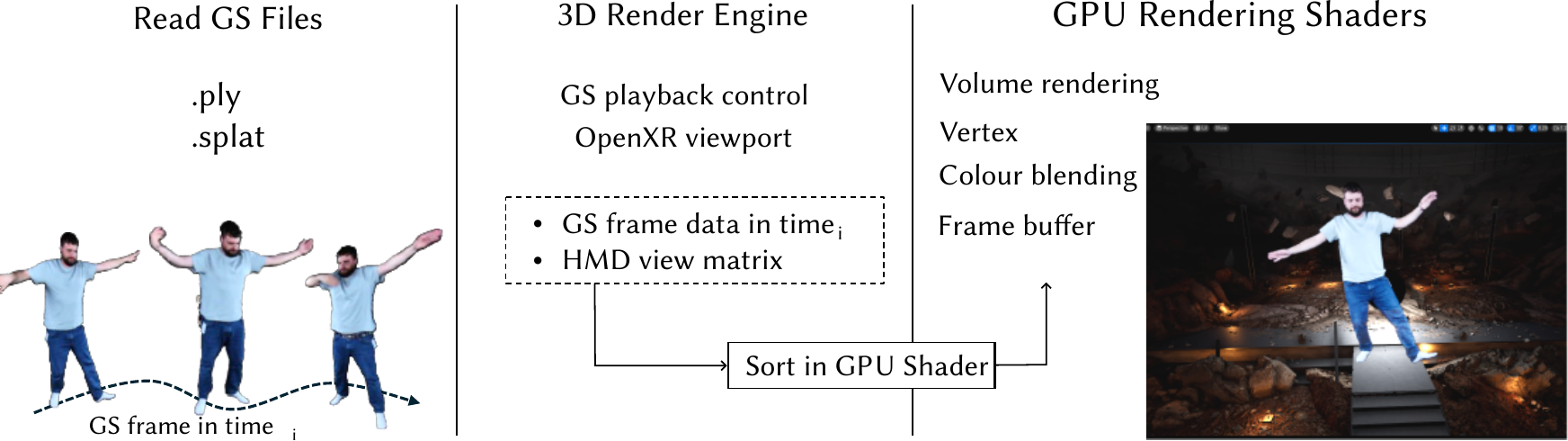}
    \caption{Gaussian Splatting (GS) Video Rendering Pipeline in 3D rendering engines.}
    \vspace{-3mm}
    \label{fig:game_eng_pipeline}
\end{figure}

For browser playback, we build a three.js/WebGL viewer \cite{threejs_gs} with WebXR support that loads \texttt{PLY}, per–frame \texttt{SPLAT}, or video \texttt{SPLAT} files and plays interactive dynamic reconstructions. Rendering uses one primitive per valid pixel and a screen–space elliptical rasterizer; to reduce aliasing we add a pixel–footprint term (\(\approx 0.3\,I_2\)) to the covariance and enforce a small minimum footprint, similar to \cite{mipsplatting}. We do not prune or build level-of-details (LODs), which keeps primitive counts moderate at our image resolutions. The viewer follows the point–cloud policy for perspective changes, switching to the per–camera Gaussian set closest to the current virtual view.

For Gaussian splat playback in VR and game engines, we additionally provide Unity and Unreal plugins. The general pipeline is given at Figure \ref{fig:game_eng_pipeline}: A 3D object is created in the engine, frame data are streamed and updated over time, and on each frame or camera move the splats are depth-sorted on the GPU. A shader then transforms splats to the current view, projects 3D covariances to 2D, rasterizes and alpha-blends elliptical colors in screen space (frame buffer) to produce the final frame. Unity and Unreal plugins will be made available upon publication.

\section{Conclusions}

We have presented a volumetric capture and reconstruction system capable of processing RGB-D or RGB-only inputs to produce point clouds and Gaussian splats. By taking the GPS-Gaussian regressor and improving it, our system achieves reliable and efficient Gaussian splat reconstructions. Moreover, it supports in-the-wild operation under uncontrolled illumination and arbitrary backgrounds. Its flexible camera configurations and easy setup and deployment make it suitable for a wide range of capture scenarios. The reconstructions can be exported in standard formats and visualized through web-based and game engine-integrated playback. Moreover, live on-site previews enable immediate feedback. The system has been evaluated and is released open-source, providing a robust platform for further research and applications in volumetric video.

As future work, we plan to further extend and improve the system. We intend to improve both the quality and compression of Gaussian splat reconstructions by adding primitive tracking and bitrate-aware coding of splat attributes.

\balance
\begingroup
\small
\begin{acks}
This work was supported by the Swiss Innovation Agency Innosuisse (108.779 INT-ICT, FaVoRe) and the European Commission Horizon programme (GA 101135637, HEAT https://heat-xr.eu/). 
\end{acks}

\bibliographystyle{ACM-Reference-Format}
\bibliography{bibliography/refs}

@misc{vpcc,
  author       = {{ISO/IEC JTC 1/SC 29}},
  title        = {Information technology — Coded representation of immersive media — Part 5: Visual volumetric video-based coding (V3C) and video-based point cloud compression (V-PCC)},
  howpublished = {International Standard ISO/IEC 23090-5:2025},
  year         = {2025},
  url          = {https://www.iso.org/standard/89030.html},
  note         = {Third edition}
}

@article{humanrf,
  title = {HumanRF: High-Fidelity Neural Radiance Fields for Humans in Motion},
  author = {I\c{s}{\i}k, Mustafa and Rünz, Martin and Georgopoulos, Markos and Khakhulin, Taras
    and Starck, Jonathan and Agapito, Lourdes and Nießner, Matthias},
  journal = {ACM Transactions on Graphics (TOG)},
  volume = {42},
  number = {4},
  pages = {1--12},
  year = {2023},
  publisher = {ACM New York, NY, USA},
  doi = {10.1145/3592415},
  url = {https://doi.org/10.1145/3592415}
}

@inproceedings{neuraldome,
  title={Neuraldome: A neural modeling pipeline on multi-view human-object interactions},
  author={Zhang, Juze and Luo, Haimin and Yang, Hongdi and Xu, Xinru and Wu, Qianyang and Shi, Ye and Yu, Jingyi and Xu, Lan and Wang, Jingya},
  booktitle={Proceedings of the IEEE/CVF Conference on Computer Vision and Pattern Recognition},
  pages={8834--8845},
  year={2023}
}

@article{renderme,
  title={Renderme-360: A large digital asset library and benchmarks towards high-fidelity head avatars},
  author={Pan, Dongwei and Zhuo, Long and Piao, Jingtan and Luo, Huiwen and Cheng, Wei and Wang, Yuxin and Fan, Siming and Liu, Shengqi and Yang, Lei and Dai, Bo and others},
  journal={Advances in Neural Information Processing Systems},
  volume={36},
  pages={7993--8005},
  year={2023}
}

@inproceedings{4dgs,
  title={4d gaussian splatting for real-time dynamic scene rendering},
  author={Wu, Guanjun and Yi, Taoran and Fang, Jiemin and Xie, Lingxi and Zhang, Xiaopeng and Wei, Wei and Liu, Wenyu and Tian, Qi and Wang, Xinggang},
  booktitle={Proceedings of the IEEE/CVF conference on computer vision and pattern recognition},
  pages={20310--20320},
  year={2024}
}

@inproceedings{gpsgaussian,
  title={Gps-gaussian: Generalizable pixel-wise 3d gaussian splatting for real-time human novel view synthesis},
  author={Zheng, Shunyuan and Zhou, Boyao and Shao, Ruizhi and Liu, Boning and Zhang, Shengping and Nie, Liqiang and Liu, Yebin},
  booktitle={Proceedings of the IEEE/CVF conference on computer vision and pattern recognition},
  pages={19680--19690},
  year={2024}
}

@online{threejs_gs, title={GaussianSplats3D}, url={https://github.com/mkkellogg/GaussianSplats3D}, journal={GitHub}, year={2025}, month={Jan}, urldate = {2025-08-06}}

@online{supersplat, title={SuperSplat}, url={https://superspl.at/editor}, journal={Superspl.at}, year={2025}, urldate = {2025-08-06}}

@Article{3dgs,
      author       = {Kerbl, Bernhard and Kopanas, Georgios and Leimk{\"u}hler, Thomas and Drettakis, George},
      title        = {3D Gaussian Splatting for Real-Time Radiance Field Rendering},
      journal      = {ACM Transactions on Graphics},
      number       = {4},
      volume       = {42},
      month        = {July},
      year         = {2023},
      url          = {https://repo-sam.inria.fr/fungraph/3d-gaussian-splatting/}
}

@online{antimatter15_splat, title={splat}, url={https://github.com/antimatter15/splat}, journal={GitHub}, year={2025}, urldate = {Accessed: 2025-08-06} }

@techreport{webxr,
  title        = {WebXR Device API},
  author       = {{Immersive Web Working Group}},
  institution  = {World Wide Web Consortium (W3C)},
  type         = {W3C Candidate Recommendation Draft},
  year         = {2025},
  month        = {Apr},
  day          = {17},
  url          = {https://www.w3.org/TR/webxr/},
  urldate      = {2025-08-15}
}

@techreport{ply,
  author       = {Greg Turk},
  title        = {The PLY Polygon File Format},
  institution  = {Stanford University Computer Graphics Laboratory},
  year         = {1994},
  url          = {https://gamma.cs.unc.edu/POWERPLANT/papers/ply.pdf},
  note         = {Version 1.0}
}

@article{craig,
  title={Introduction to robotics: Mechanics and control},
  author={Craig, John J.},
  journal={IEEE Journal on Robotics and Automation},
  volume={3},
  number={2},
  pages={166--166},
  year={1987},
  publisher={IEEE}
}

@article{kdtree,
  title={Multidimensional binary search trees used for associative searching},
  author={Bentley, Jon Louis},
  journal={Communications of the ACM},
  volume={18},
  number={9},
  pages={509--517},
  year={1975},
  publisher={ACM New York, NY, USA}
}

@inproceedings{foundationstereo,
  title={Foundationstereo: Zero-shot stereo matching},
  author={Wen, Bowen and Trepte, Matthew and Aribido, Joseph and Kautz, Jan and Gallo, Orazio and Birchfield, Stan},
  booktitle={Proceedings of the Computer Vision and Pattern Recognition Conference},
  pages={5249--5260},
  year={2025}
}

@inproceedings{raftstereo,
  title={Raft-stereo: Multilevel recurrent field transforms for stereo matching},
  author={Lipson, Lahav and Teed, Zachary and Deng, Jia},
  booktitle={2021 International Conference on 3D Vision (3DV)},
  pages={218--227},
  year={2021},
  organization={IEEE}
}

@inproceedings{thuman,
  title={Function4d: Real-time human volumetric capture from very sparse consumer rgbd sensors},
  author={Yu, Tao and Zheng, Zerong and Guo, Kaiwen and Liu, Pengpeng and Dai, Qionghai and Liu, Yebin},
  booktitle={Proceedings of the IEEE/CVF conference on computer vision and pattern recognition},
  pages={5746--5756},
  year={2021}
}

@inproceedings{mipsplatting,
  title={Mip-splatting: Alias-free 3d gaussian splatting},
  author={Yu, Zehao and Chen, Anpei and Huang, Binbin and Sattler, Torsten and Geiger, Andreas},
  booktitle={Proceedings of the IEEE/CVF conference on computer vision and pattern recognition},
  pages={19447--19456},
  year={2024}
}

@article{friedman1977algorithm,
  title={An algorithm for finding best matches in logarithmic expected time},
  author={Friedman, Jerome H. and Bentley, Jon Louis and Finkel, Raphael Ari},
  journal={ACM Transactions on Mathematical Software (TOMS)},
  volume={3},
  number={3},
  pages={209--226},
  year={1977},
  publisher={ACM}
}

@article{ssim,
  title={Image quality assessment: from error visibility to structural similarity},
  author={Wang, Zhou and Bovik, Alan C and Sheikh, Hamid R and Simoncelli, Eero P},
  journal={IEEE Transactions on Image Processing},
  volume={13},
  number={4},
  pages={600--612},
  year={2004},
  publisher={IEEE}
}

@inproceedings{tomasi1998bilateral,
  author    = {Carlo Tomasi and Roberto Manduchi},
  title     = {Bilateral Filtering for Gray and Color Images},
  booktitle = {Proceedings of the IEEE International Conference on Computer Vision (ICCV)},
  year      = {1998},
  pages     = {839--846},
  doi       = {10.1109/ICCV.1998.710815}
}

@online{caliscope,
  title  = {Caliscope},
  year   = {2025},
  url    = {https://github.com/mprib/caliscope},
  urldate = {2025-08-06}
}

@online{multicamcalib,
  title  = {MultiCamCalib},
  year   = {2025},
  url    = {https://github.com/hjoonpark/MultiCamCalib},
  urldate = {2025-08-06}
}

@online{celery,
  title  = {Celery},
  year   = {2025},
  url    = {https://docs.celeryq.dev/en/stable/index.html},
  urldate = {2025-08-06}
}

@inproceedings{sterzentsenko2018low,
  title={A low-cost, flexible and portable volumetric capturing system},
  author={Sterzentsenko, Vladimiros and Karakottas, Antonis and Papachristou, Alexandros and Zioulis, Nikolaos and Doumanoglou, Alexandros and Zarpalas, Dimitrios and Daras, Petros},
  booktitle={2018 14th International Conference on Signal-Image Technology \& Internet-Based Systems (SITIS)},
  pages={200--207},
  year={2018},
  organization={IEEE}
}

@online{volumetricapture,
  title  = {Volumetric Capture},
  year   = {2025},
  url    = {https://vcl3d.github.io/VolumetricCapture/},
  xnote   = {Accessed: 2025-08-06}
}

@online{depthkitstudio,
  title  = {DepthKit Studio},
  year   = {2025},
  url    = {https://www.depthkit.tv/depthkit-studio},
  xnote   = {Accessed: 2025-08-06}
}

@inproceedings{kowalski2015livescan3d,
  title={Livescan3d: A fast and inexpensive 3d data acquisition system for multiple kinect v2 sensors},
  author={Kowalski, Marek and Naruniec, Jacek and Daniluk, Michal},
  booktitle={2015 international conference on 3D vision},
  pages={318--325},
  year={2015},
  organization={IEEE}
}

@online{livescan3d,
  title  = {LiveScan3D},
  year   = {2025},
  url    = {https://github.com/MarekKowalski/LiveScan3D},
  xnote   = {Accessed: 2025-08-06}
}

@online{scannedreality,
  title  = {Scanned Reality Studio},
  year   = {2025},
  url    = {https://scanned-reality.com/},
  xnote   = {Accessed: 2025-08-06}
}

@online{evercoast,
  title  = {Evercoast},
  year   = {2025},
  url    = {https://www.evercoast.com/},
  xnote   = {Accessed: 2025-08-06}
}

@online{4dviews,
  title  = {4Dviews},
  year   = {2025},
  url    = {https://www.4dviews.com/},
  xnote   = {Accessed: 2025-08-06}
}

@online{volograms,
  title  = {Volograms},
  year   = {2025},
  url    = {https://www.volograms.com/},
  xnote   = {Accessed: 2025-08-06}
}

@online{8i,
  title  = {8i},
  year   = {2025},
  url    = {https://8i.com/},
  xnote   = {Accessed: 2025-08-06}
}

@incollection{xu2023easyvolcap,
  title={Easyvolcap: Accelerating neural volumetric video research},
  author={Xu, Zhen and Xie, Tao and Peng, Sida and Lin, Haotong and Shuai, Qing and Yu, Zhiyuan and He, Guangzhao and Sun, Jiaming and Bao, Hujun and Zhou, Xiaowei},
  booktitle={SIGGRAPH Asia 2023 Technical Communications},
  pages={1--4},
  year={2023}
}

@online{volucap,
  title  = {Volucap},
  year   = {2025},
  url    = {https://volucap.com/},
  xnote   = {Accessed: 2025-08-06}
}

@online{microsoft,
  title  = {Microsoft mixed reality capture studios},
  url    = {https://news.microsoft.com/source/features/work-life/microsoft-mixed-reality-capture-studios-create-holograms-to-educate-and-entertain/},
  year    = {2025},
  xnote   = {Accessed: 2025-08-06}
}

@inproceedings{yang2025imvid,
  title={ImViD: Immersive Volumetric Videos for Enhanced VR Engagement},
  author={Yang, Zhengxian and Pan, Shi and Wang, Shengqi and Wang, Haoxiang and Lin, Li and Li, Guanjun and Wen, Zhengqi and Lin, Borong and Tao, Jianhua and Yu, Tao},
  booktitle={Proceedings of the Computer Vision and Pattern Recognition Conference},
  pages={16554--16564},
  year={2025}
}

@article{Collet2015,
	title        = {High-quality streamable free-viewpoint video},
	author       = {Collet, Alvaro and Chuang, Ming and Sweeney, Pat and Gillett, Don and Evseev, Dennis and Calabrese, David and Hoppe, Hugues and Kirk, Adam and Sullivan, Steve},
	year         = 2015,
	journal      = {ACM Transactions on Graphics (ToG)},
	publisher    = {ACM New York, NY, USA},
	volume       = 34,
	number       = 4,
	pages        = {1--13}
}

@inproceedings{orts2016holoportation,
  title={Holoportation: Virtual 3d teleportation in real-time},
  author={Orts-Escolano, Sergio and Rhemann, Christoph and Fanello, Sean and Chang, Wayne and Kowdle, Adarsh and Degtyarev, Yury and Kim, David and Davidson, Philip L and Khamis, Sameh and Dou, Mingsong and others},
  booktitle={Proceedings of the 29th annual symposium on user interface software and technology},
  pages={741--754},
  year={2016}
}

@inproceedings{schreer2019capture,
  title={Capture and 3D video processing of volumetric video},
  author={Schreer, Oliver and Feldmann, Ingo and Renault, Sylvain and Zepp, Marcus and Worchel, Markus and Eisert, Peter and Kauff, Peter},
  booktitle={2019 IEEE International conference on image processing (ICIP)},
  pages={4310--4314},
  year={2019},
  organization={IEEE}
}

@software{yolo11,
  author = {Glenn Jocher and Jing Qiu},
  title = {Ultralytics YOLO11},
  version = {11.0.0},
  year = {2024},
  url = {https://github.com/ultralytics/ultralytics},
  orcid = {0000-0001-5950-6979, 0000-0003-3783-7069},
  license = {AGPL-3.0}
}

@article{sam2,
  title={SAM 2: Segment Anything in Images and Videos},
  author={Ravi, Nikhila and Gabeur, Valentin and Hu, Yuan-Ting and Hu, Ronghang and Ryali, Chaitanya and Ma, Tengyu and Khedr, Haitham and R{\"a}dle, Roman and Rolland, Chloe and Gustafson, Laura and Mintun, Eric and Pan, Junting and Alwala, Kalyan Vasudev and Carion, Nicolas and Wu, Chao-Yuan and Girshick, Ross and Doll{\'a}r, Piotr and Feichtenhofer, Christoph},
  journal={arXiv preprint arXiv:2408.00714},
  url={https://arxiv.org/abs/2408.00714},
  year={2024}
}

@inproceedings{videoflow,
  title={Videoflow: Exploiting temporal cues for multi-frame optical flow estimation},
  author={Shi, Xiaoyu and Huang, Zhaoyang and Bian, Weikang and Li, Dasong and Zhang, Manyuan and Cheung, Ka Chun and See, Simon and Qin, Hongwei and Dai, Jifeng and Li, Hongsheng},
  booktitle={Proceedings of the IEEE/CVF International Conference on Computer Vision},
  pages={12469--12480},
  year={2023}
}

@inproceedings{teed2020raft,
  title={Raft: Recurrent all-pairs field transforms for optical flow},
  author={Teed, Zachary and Deng, Jia},
  booktitle={European conference on computer vision},
  pages={402--419},
  year={2020},
  organization={Springer}
}

@inproceedings{flowformer,
  title={Flowformer: A transformer architecture for optical flow},
  author={Huang, Zhaoyang and Shi, Xiaoyu and Zhang, Chao and Wang, Qiang and Cheung, Ka Chun and Qin, Hongwei and Dai, Jifeng and Li, Hongsheng},
  booktitle={European conference on computer vision},
  pages={668--685},
  year={2022},
  organization={Springer}
}

@InProceedings{4dgs1,
    author    = {Wu, Guanjun and Yi, Taoran and Fang, Jiemin and Xie, Lingxi and Zhang, Xiaopeng and Wei, Wei and Liu, Wenyu and Tian, Qi and Wang, Xinggang},
    title     = {4D Gaussian Splatting for Real-Time Dynamic Scene Rendering},
    booktitle = {Proceedings of the IEEE/CVF Conference on Computer Vision and Pattern Recognition (CVPR)},
    month     = {June},
    year      = {2024},
    pages     = {20310-20320}
}

@inproceedings{4dgs2,
  author = {Fang, Jiemin and Yi, Taoran and Wang, Xinggang and Xie, Lingxi and Zhang, Xiaopeng and Liu, Wenyu and Nie\ss{}ner, Matthias and Tian, Qi},
  title = {Fast Dynamic Radiance Fields with Time-Aware Neural Voxels},
  year = {2022},
  booktitle = {SIGGRAPH Asia 2022 Conference Papers}
}
\endgroup

\appendix

\end{document}